\documentclass[12pt]{article}
\usepackage{a4wide}
\usepackage{amsmath,amssymb}
\usepackage{latexsym,graphicx,epsfig,cite}

\def\beq{\begin{equation}}
\def\eeq{\end{equation}}
\def\bea{\begin{eqnarray}}
\def\eea{\end{eqnarray}}

\def\lsim{\mathrel{\rlap{\raise 2.5pt \hbox{$<$}}\lower 2.5pt\hbox{$\sim$}}}
\def\gsim{\mathrel{\rlap{\raise 2.5pt \hbox{$>$}}\lower 2.5pt\hbox{$\sim$}}}

\newcommand{\half}{{\textstyle\frac{1}{2}}}

\newcommand{\pslash}{\rlap/p}
\newcommand{\qslash}{\rlap/q}
\newcommand{\kslash}{\rlap/k}
\newcommand{\epsilonslash}{\rlap/\epsilon}

\newcommand{\slepton}{{\tilde\ell}}
\newcommand{\gravitino}{{\tilde G}}

\renewcommand{\Re}{{\rm Re\thinspace}}

\allowdisplaybreaks %!!!!
\begin{document}

\pagestyle{empty}
\begin{flushright}
{CERN-PH-TH/2007-122}
\end{flushright}
\vspace*{5mm}
\begin{center}
{\large {\bf Radiative gravitino decays from R-parity violation}} \\
\vspace*{1cm}
{\bf S.\ Lola$^1$, P.\ Osland$^2$} and {\bf A.R.\ Raklev$^2$} \\
\vspace{0.3cm}
$^1$ Department of Physics, University of Patras, GR-26500 Patras, Greece \\
$^2$ Department of Physics and Technology, University of Bergen, 
Postboks 7803, \\
N-5020 Bergen, Norway

\vspace*{2cm}
{\bf ABSTRACT}
\end{center}
\vspace*{2mm}

We study radiative gravitino decay within the framework of R-violating
supersymmetry.  For trilinear R-violating couplings that involve the
third generation of fermions, or for light gravitinos, we find that
the radiative loop-decay $\tilde{G} \rightarrow \gamma \nu$ dominates
over the tree-level ones for a wide set of parameters. We calculate
the gravitino decay width and study its implications for cosmology and
collider physics. Slow-decaying gravitinos are good dark matter
candidates, for a range of parameters that would also predict
observable R-violating signatures in colliders. In general the
branching ratios are very dependent on the relative hierarchies of
R-violating operators, and may provide relevant information on the
flavour structure of the underlying fundamental theory.

%\vspace*{2cm}

%\vfill\eject

\setcounter{page}{1}
\pagestyle{plain}

\pagebreak

\section{Introduction}

The possibility of R-violating supersymmetry \cite{Rpar,Hall:1983id}
has been extensively studied, as an alternative scenario to the
Minimal Supersymmetric Standard Model (MSSM) \cite{HabKan}. Indeed,
the symmetries of the Standard Model, in addition to the Yukawa
couplings that generate fermion masses, allow for additional
dimension-four couplings of the form
\begin{equation}
\lambda L_{i}L_{j}{\bar{E}}_{k}+\lambda ^{\prime }L_{i}Q_{j}{\bar{D}_{k}}
+\lambda ^{\prime \prime }{\bar{U}_{i}}{\bar{D}_{j}}{\bar{D}_{k}} 
\end{equation}
where the $L(Q)$ are the left-handed lepton (quark) doublet
superfields, and the ${\bar{E}}$ (${\bar{D}},{\bar{U}}$) are the
corresponding left-handed singlet fields. Amongst these so-called
R-violating couplings, the first two violate lepton number, while the
third violates baryon number, and the symmetries of the theory imply
that there are 45 operators in total --- 9, 27, and 9 respectively for
the three terms, due to $SU(2)$ and $SU(3)$ invariance.

The stricter bounds on R-violating operators come from proton
stability.  However, it is by now well-understood that R-parity
\cite{fayet} is not the only symmetry that can guarantee proton
stability; baryon or lepton parities \cite{IR,LR} can have the same
effect.  In fact, it is crucial to exclude the simultaneous presence
of only certain products of $LQ\bar{D}$ and $\bar{U}\bar{D}\bar{D}$
couplings~\cite{SMVIS}.  In addition to proton stability, most
operators are subject to experimental constraints from the
non-observation of modifications to Standard Model processes, or of
possible exotic processes~\cite{constraints}\footnote{Additional
constraints can be obtained from a possible detection of photon or
lepton plus missing energy signatures from NLSP decays to a gravitino
LSP \cite{CPW}.}.

Going beyond proton stability and parameter constraints, one
of the reasons that R-violating supersymmetry has not seemed very
appealing, is the notion that in these models the lightest
supersymmetric particle (LSP) is not stable, and therefore one has to
search elsewhere for dark matter.

However, how absolute is this statement?  Is it possible that the
lightest supersymmetric particle decays so slowly that {\em
cosmologically it is almost stable}, while at the same time there is
at least one coupling large enough to lead to observable collider
signatures for R-violation? In this letter we shall take this to mean
that at least one coupling is larger than approximately $10^{-6}$ for
sparticles of 100 GeV mass. This could for instance happen, if the LSP
is a light gravitino whose R-parity violating decays to conventional
particles are for some reason very suppressed. Scenarios with light
gravitinos are well-motivated --- and one could even expect masses as
low as $10^{-5}$ eV \cite{Zw}.

Tree-level R-violating gravitino decays have been studied in
\cite{CM},
%
%where the focus was on schemes where the
%gravitino and scalar superparticle masses do not exceed {\cal O}(10~TeV), and 
%
where gravitinos were found to be able to decay before the present
epoch for large values of the R-violating couplings and gravitino
masses, but not earlier than the start of big-bang nucleosynthesis
(BBN). It was concluded that the considered scenario would upset the
standard BBN scenario, and did not seem to constitute a natural
solution for the cosmological gravitino problem. Two-body gravitino
decays to photon and neutrino, for bilinear R-parity violation, were
studied in \cite{TY} --- with the decays arising from the mixing of
neutralinos with neutrinos --- and it was concluded that gravitinos of
less than 1 GeV mass can be the dark matter and at the same time
consistent with neutrino mass generation from bilinear R-violation.
Recently, in \cite{BM}, these decays were revisited, and in models
where R-parity breaking is tied to B-L breaking, very small
R-violating couplings were predicted, in conjunction with a photon
flux that can explain the apparent EGRET excess in the extragalactic
diffuse gamma-ray flux \cite{reEGRET} for $\sim10$~GeV gravitinos. In
addition, radiative neutralino decays in R-violating schemes have been
studied in \cite{Hall:1983id,rad-neut}.

Here we will focus on radiative gravitino decays 
\begin{equation} \label{Eq:process}
\tilde G\to\gamma\nu,
\end{equation}
in schemes with explicitly broken R-violation from trilinear terms,
and will compare them to the tree-level three-body decays that are
expected from the same terms. Which processes will dominate, and what
will be the cosmological and collider signatures depends on the
flavour structure of the R-violating operators, which we will also
discuss.

In principle one may imagine that gravitinos are almost-stable and
could ``act'' as dark matter in the following cases:

\begin{itemize}
\item[$\bullet$]
Suppose the only relevant operators are the $\lambda ^{\prime \prime }
\bar{U}_{3} \bar{D}_j \bar{D}_k$. In the three-body gravitino decays
discussed in \cite{CM} there would inevitably be a top-quark in the
final state. If the gravitino is lighter than the top quark --- even
if heavier than all other fermions --- then it is stable with respect
to tree-body decays, up to mixing effects\footnote{Note that for an
operator of the form $\lambda ^{\prime} L_i Q_3 \bar{D}_k$ this
argument does not hold, since, when we pass from superfields to
component fields the $L_iQ_3$ part can become $\ell_i t$ or 
$\nu_i b$.}.

\item[$\bullet$]
This brings us to the next step: Suppose that all dominant operators
involve third-generation fermions, and lead to bottom-quark or tau
final states. For a gravitino below the $\sim 1$~GeV scale, tree-level
decays will again be very suppressed.

\item[$\bullet$]
Finally, for ``super-light'' gravitinos, as in \cite{Zw}, gravitinos
are essentially stable under the 3-body decays!
\end{itemize}

In either of the above cases --- classified according to the possible
range of gravitino masses for the suppression of three-body decays ---
gravitinos have a very large lifetime, that can exceed the age of the
universe. In the case of a $\lambda ^{\prime \prime } \bar{U}_{3}
\bar{D}_j \bar{D}_k$ operator, tree-level decays can proceed only via
quark mixing (which in some models could be very suppressed). For
$LL{\bar{E}}$ or $LQ{\bar{D}}$ operators, the radiative decay of the
gravitino can still dominate over the tree-body modes. In particular
for the cases where third-generation fermions are involved, where both
the loop factor of the radiative mode, and the phase-space suppression
of the tree-level diagrams are larger.

In what follows, we will calculate the amplitude for radiative
gravitino decays and discuss the effects of the underlying flavour
structure in Section~\ref{sec:rad}, leading to a comparison between
radiative and tree-level gravitino decays in
Section~\ref{sec:radvs3b}. We outline the related cosmological
implications for dark matter, photon spectra and big-bang
nucleosynthesis, and discuss possible collider phenomenology from the
decays of the next-to-lightest supersymmetric particle (NLSP) in
Section~\ref{sec:cosmocoll} that could help to further distinguish
between the different scenarios of slowly decaying gravitinos, before
we Conclude.

%%%%%%%%%%%%%%%%%%%%%%%%%%%%%%%%%%%%%%%%%%%%%%%%%%%%%%%%%%%%%%%%%%%%%%%%
\section{Radiative gravitino decays}
\label{sec:rad}
\setcounter{equation}{0}
%%%%%%%%%%%%%%%%%%%%%%%%%%%%%%%%%%%%%%%%%%%%%%%%%%%%%%%%%%%%%%%%%%%%%%%%

There are three main structures of diagrams that generate the decay
(\ref{Eq:process}), as shown in Fig.~\ref{Fig:feyn}. Note that the
class of operators that can be involved are of the form
$LL_j\bar{E}_j$ or $LQ_j\bar{D}_j$. In the latter case the neutrino is
coupled to down-type quark to preserve SU(2) invariance. The common
$j$ index appears since the same fermion flavour should couple to both
the gravitino and the photon. In Fig.~\ref{Fig:3-body} we show the
corresponding three-body decay induced by the same operators.

%%%%%%%%%%%%%%%%%%%%%%%%%%%%%%%%%%%%%%%%%%%%%%%%%%%%%%%%%%%%%%%%%%%%%%%%
\begin{figure}[htb]
\refstepcounter{figure}
\label{Fig:feyn}
\addtocounter{figure}{-1}
\begin{center}
\setlength{\unitlength}{1cm}
\begin{picture}(15.0,7.0)
\put(0.,0.)
{\mbox{\epsfysize=7cm
 \epsffile{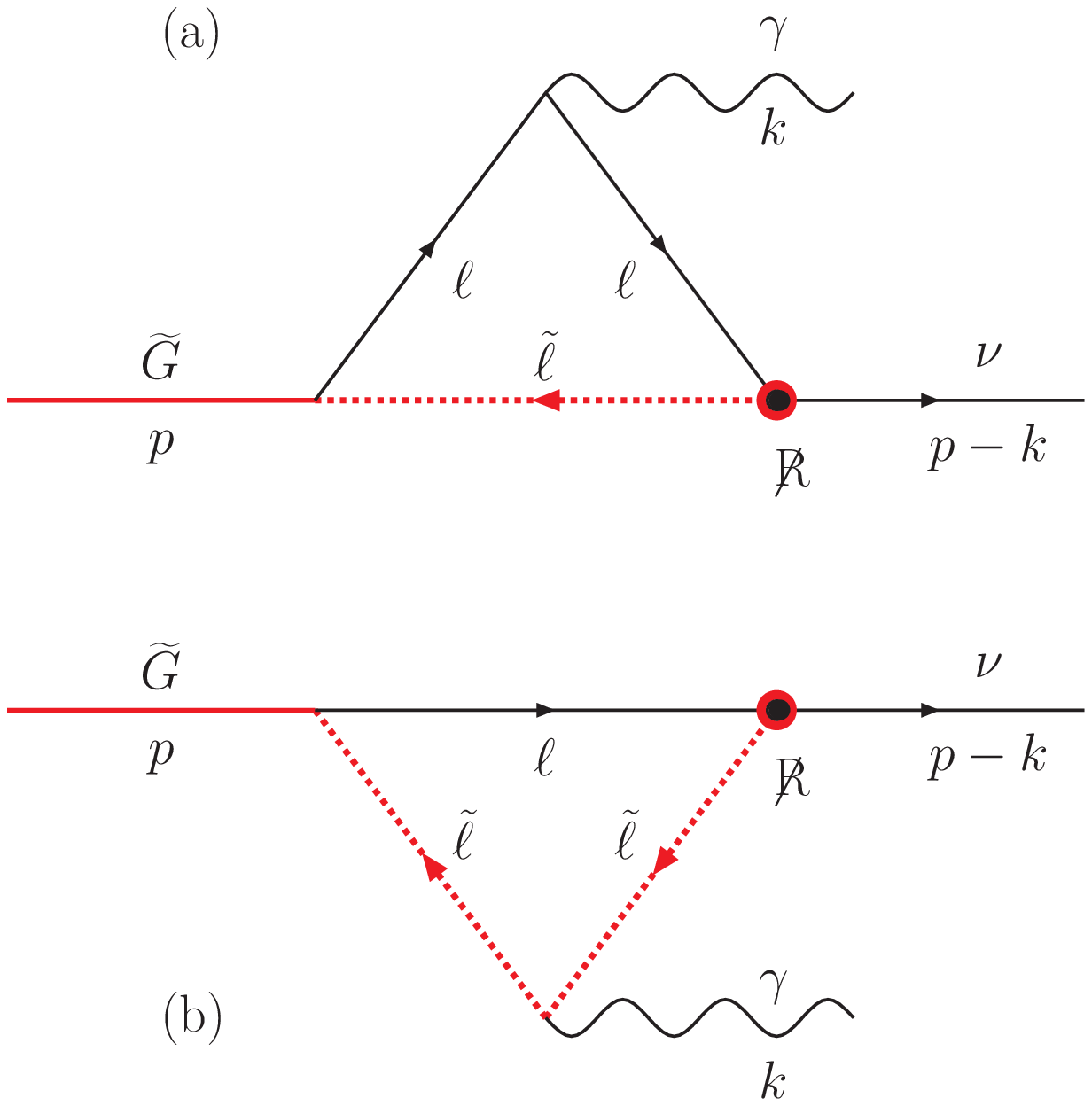}}}
\put(7.5,1.7)
{\mbox{\epsfysize=3.0cm
 \epsffile{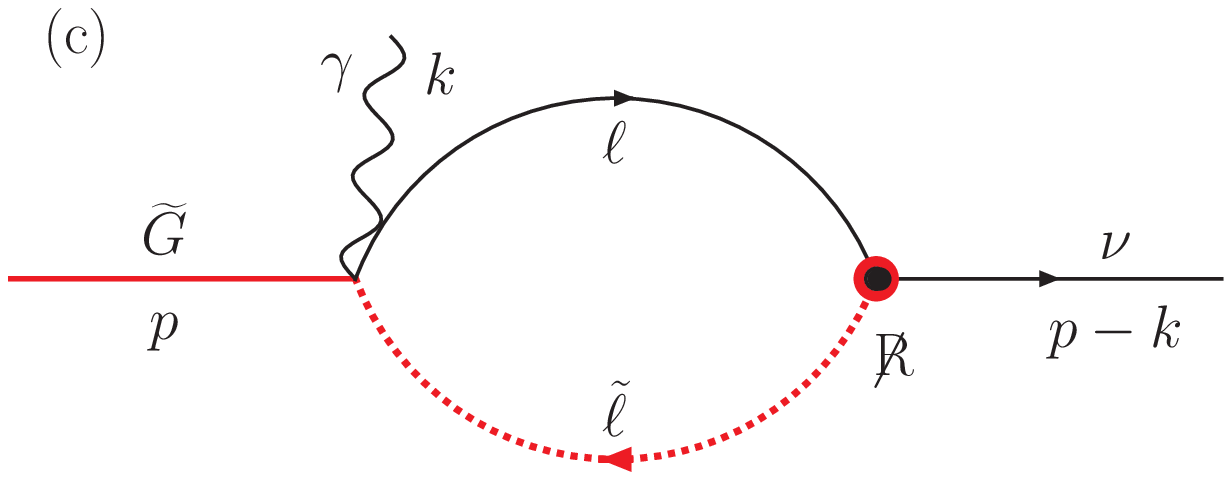}}}
\end{picture}
%\vspace*{-4mm}
\caption{Basic set of Feynman diagrams for radiative gravitino decay,
shown for (s)lepton loops. Arrows denote flow of lepton number for
left-chiral fields.}
\end{center}
\end{figure}
%%%%%%%%%%%%%%%%%%%%%%%%%%%%%%%%%%%%%%%%%%%%%%%%%%%%%%%%%%%%%%%%%%%%%%

%%%%%%%%%%%%%%%%%%%%%%%%%%%%%%%%%%%%%%%%%%%%%%%%%%%%%%%%%%%%%%%%%%%%%%%%
\begin{figure}[htb]
\refstepcounter{figure}
\label{Fig:3-body}
\addtocounter{figure}{-1}
\begin{center}
\setlength{\unitlength}{1cm}
\begin{picture}(15.0,4.5)
\put(3.5,0.)
{\mbox{\epsfysize=5cm
 \epsffile{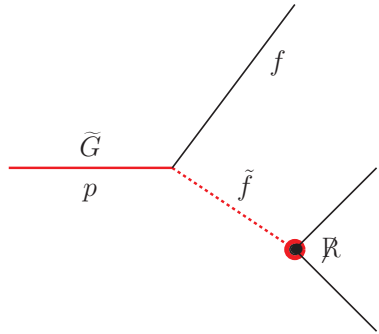}}}
\end{picture}
%\vspace*{-4mm}
\caption{Three-body decay of gravitino via R-parity violating coupling.}
\end{center}
\end{figure}
%%%%%%%%%%%%%%%%%%%%%%%%%%%%%%%%%%%%%%%%%%%%%%%%%%%%%%%%%%%%%%%%%%%%%%

The $R$-parity violating coupling for incoming left-chiral fields can
be parametrised as\footnote{For outgoing left-chiral fields, this
becomes $+i\lambda P_R$.}
\begin{equation} \label{Eq:R-parity-violation}
\tilde\ell\ell'\nu_\ell:  \quad -i\lambda P_L,
\end{equation}
with $\lambda$ a dimensionless quantity, and
$P_{L/R}=\half(1\mp\gamma_5)$.  The photonic coupling to sparticles is
given by
\begin{equation}
\slepton^+\slepton^-A^\nu:
\quad +ie(q_+-q_-)^\nu,
\end{equation}
with $q_{-}$ and $q_{+}$ the incoming slepton momenta.
The gravitino couplings are (see \cite{Moroi:1995fs,Bolz:2000xi}):
\begin{alignat}{2} \label{Eq:Bolz}
&\gravitino^\mu\tilde\ell\ell:  &\quad 
&\frac{i}{\sqrt{2}M}\, P_R\,\gamma^\mu\qslash_{-}, \nonumber \\
&\gravitino^\mu\tilde\ell\ell A^\nu:  &\quad 
&\frac{-ie}{\sqrt2 M}P_R \gamma^\mu\gamma^\nu,
\end{alignat}
with $M=(8\pi G_N)^{-1/2}=2.4\times10^{18}~\text{GeV}$ the reduced
Planck mass.

In order to have a non-zero contribution, there must be an even number
of $\gamma$ matrices between the projection operators at the vertices
where the gravitino and the neutrino couple. Thus, we immediately see
that the amplitude is proportional to the lepton mass, $m_\ell$ --- or
a down-type quark mass, for the $LQ\bar{D}$ case. This amounts to a
mass term insertion, as is required in order to get the correct
helicity combinations at each vertex. In addition to the diagrams
shown in Fig.~\ref{Fig:feyn} there is a corresponding set of diagrams
with reversed fermion flow and right-handed (s)leptons.

%%%%%%%%%%%%%%%%%%%%%%%%%%%%%%%%%%%%%%%%%%%%%%%%%%%%%%%%%%%%%%%%%%%%%%%%
\subsection{Feynman amplitudes and gauge invariance}
%%%%%%%%%%%%%%%%%%%%%%%%%%%%%%%%%%%%%%%%%%%%%%%%%%%%%%%%%%%%%%%%%%%%%%%%

Extracting one lepton mass factor, the total amplitude for the Feynman
diagrams shown in Fig.~\ref{Fig:feyn} can be written as
\begin{equation} \label{Eq:calM}
{\cal M}=-\frac{ie\lambda\,m_\ell}{16\sqrt{2}\pi^2\,M}\,
m_\gravitino\,{\cal F}(m_\gravitino^2, m_\slepton^2, m_\ell^2).
\end{equation}
Using the equations of motion of the gravitino \cite{Moroi:1995fs}:
\begin{equation} \label{eq:eq-motion-gravitino}
\gamma_\mu\tilde\psi^\mu(p)=0, \quad
p_\mu\tilde\psi^\mu(p)=0, \quad
(\pslash-m_\gravitino)\tilde\psi^\mu(p)=0,
\end{equation}
and of the neutrino: $\pslash\, u(p-k)=\kslash\, u(p-k)$ (neglecting
the neutrino mass), one can simplify the matrix elements
considerably. The dimensionless quantity ${\cal F}$ can thus be
written as
\begin{equation} \label{Eq:calF}
{\cal F}=
\bar u(p-k)\bigl[a\, k_\mu\,\epsilonslash+b\, k_\mu(\epsilon\cdot p)
+c\,\epsilon_\mu \bigr]
\tilde\psi^\mu(p).
\end{equation}
The constraint of gauge invariance imposes a relation on $a$, $b$ and
$c$, and we are left with {\it two} gauge-independent structures.
This is different from the case of $\tilde\chi_0\to\gamma\nu$
\cite{Hall:1983id,rad-neut} where there is only one amplitude.

Using the kinematical relation, $(p-k)^2=0$, or $2p\cdot
k=m_\gravitino^2$, it is convenient to choose the two amplitudes as
follows:
\begin{equation} \label{Eq:calF-gauge}
{\cal F}=
\bar u(p-k)P_R\bigl[
c_1{\cal A}_\mu^{(1)}+c_2{\cal A}_\mu^{(2)}\bigr]\tilde\psi^\mu(p),
\end{equation}
where $c_1$ and $c_2$ are dimensionless coefficients, the overall
factor $P_R$ arises from the coupling to the left-handed neutrino, and
the structures ${\cal A}_\mu^{(1)}$ and ${\cal A}_\mu^{(2)}$ are given
by
\begin{align} \label{Eq:cal_A-structures}
{\cal A}_\mu^{(1)}&=k_\mu[2(\epsilon\cdot p)-\epsilonslash\,m_\gravitino]/
m_\gravitino^3, \nonumber \\
{\cal A}_\mu^{(2)}&=[2k_\mu(\epsilon\cdot p)-\epsilon_\mu\,m_\gravitino^2]/
m_\gravitino^3.
\end{align}

The integrals from diagrams (a) and (b) will be logarithmically
divergent and effectively regularised by the contributions of diagram
(c). Thus, we see that ${\cal F}$ can be expressed in terms of two
three-point functions, and finite differences of two-point functions.
These terms must separately be gauge independent:
\begin{equation} \label{Eq:calF-other}
{\cal F}=
\bar u(p-k)\bigl[
X_a C_0^{(a)}
+X_b C_0^{(b)}
+Y^{(1)} \Delta B_0^{(1)}
+Y^{(2)}\Delta B_0^{(2)}
\bigr]\tilde\psi^\mu(p),
\end{equation}
where the coefficients $X$ and $Y$ must be of the forms given in
(\ref{Eq:cal_A-structures}), and the $C_0$ are three-point functions
corresponding to diagrams (a) and (b), whereas the $\Delta B_0$ are finite
differences of two-point functions.  In the notation of {\tt LoopTools}
\cite{Hahn:1998yk,vanOldenborgh:1989wn}, we have
\begin{align} \label{Eq:finite_expressions}
C_0^{(a)}&=C_0(m_\gravitino^2,0,0,m_\slepton^2,m_\ell^2,m_\ell^2), \nonumber \\
C_0^{(b)}&=C_0(m_\gravitino^2,0,0,m_\slepton^2,m_\ell^2,m_\slepton^2), 
\nonumber \\
\Delta B_0^{(1)}&=2B_0(m_\gravitino^2,m_\slepton^2,m_\ell^2)
-B_0(0,m_\slepton^2,m_\ell^2)
-B_0(0,m_\ell^2,m_\ell^2), \nonumber \\
\Delta B_0^{(2)}&=B_0(m_\gravitino^2,m_\slepton^2,m_\ell^2)
-B_0(0,m_\slepton^2,m_\ell^2).
\end{align}

Adding the contributions of the three diagrams, we find that the
coefficients of the amplitude are given by:
\begin{align}
c_1&=2[(m_\gravitino^2-m_\slepton^2+m_\ell^2)C_0^{(a)}
+2B_0(m_\gravitino^2,m_\slepton^2,m_\ell^2)-B_0(0,m_\slepton^2,m_\ell^2)
-B_0(0,m_\ell^2,m_\ell^2)], \nonumber \\
c_2&=2
[m_\ell^2 C_0^{(a)}+m_\slepton^2\,C_0^{(b)}
+B_0(m_\gravitino^2,m_\slepton^2,m_\ell^2) 
-B_0(0,m_\slepton^2,m_\ell^2)].
\label{eq:c1c2}
\end{align}
These are both UV-finite, and have the structures given in
eqs.~(\ref{Eq:calF-other}) and (\ref{Eq:finite_expressions}).

%%%%%%%%%%%%%%%%%%%%%%%%%%%%%%%%%%%%%%%%%%%%%%%%%%%%%%%%%%%%%%%%%%%%%%%%
\subsection{Flavour considerations}
%%%%%%%%%%%%%%%%%%%%%%%%%%%%%%%%%%%%%%%%%%%%%%%%%%%%%%%%%%%%%%%%%%%%%%%%

Before comparing the radiative and the three-body decays, we would
like to comment on the dependence of our results on flavour physics.
The relative magnitude of radiative gravitino decays as compared to
the tree-level ones depends on the flavour structure of the
R-violating operators --- as we shall see, for higher generations the
radiative decay becomes larger and the same is true for the
suppression of the tree-level diagrams. This picture is not unknown to
us: the Yukawa couplings that generate fermion masses also have clear
hierarchies, and progressively lead to higher masses as we pass from
the first to the third generation of quarks and leptons.

Given that the same fermion fields that enter into R-violating
operators, also enter in the Yukawa mass terms, one may try to
directly link R-violating hierarchies to those of fermion masses
\cite{MODELS,ELR}. This is done using models with family symmetries.
Fermion generation charges are chosen in such a way that only the
third generation mass terms have a zero charge (and thus are allowed
when the symmetry is exact). The rest of the masses are generated at a
higher order by the spontaneous breaking of this symmetry by the
vacuum expectation values of singlet fields, and are suppressed by the
heavy mass scales of the theory \cite{Froggatt:1978nt,Ibanez:1994ig}.
If $R$ parity is violated in such models, couplings with different
family charges will also appear with different powers of the family
symmetry-breaking parameter, and thus with different magnitudes.

In general, for models appearing in the literature, the relative
flavour charges and thus the mass matrices, are determined by the GUT
multiplet structure; particles in the same GUT multiplet have the same
charge.  Nevertheless, in all cases, the observed fermion hierarchies
require smaller charges for the operators of the higher generations
--- typically zero for the top Yukawa mass terms, but also for the
bottom and tau in supersymmetric models with large $\tan\beta$. This
implies that it is reasonable to expect that R-violating operators
that contain fields of the third generation may also be
larger. Whether this happens will depend on whether fermion hierarchies can
be directly linked to the R-violating ones, or whether extra singlet
fields with non-zero flavour charge are involved, thus reversing the
hierarchies with respect to those of the masses \cite{ELR}. In this latter
case, tree-level gravitino decays are more likely to dominate, unless
gravitinos are very light.

Moreover, one should appropriately take into account mixing
effects. Indeed, {\em even in the case of one dominant operator in the
basis of current eigenstates, mixing effects will induce non-zero
coefficients for related operators in the basis of mass eigenstates}.
These will be suppressed by the mixing parameters with respect to the
dominant operator, but will be non-zero, and this may affect
phenomenological and cosmological predictions\footnote{The fact that
there are strict bounds on some operators, implies that mixing effects
may in given models generate additional bounds on couplings that at a
first glance look less constrained. This has been analysed in detail
in \cite{ELR}, where it was shown that in theories with strong
correlations between operators (such as left-right symmetric models),
the effects can be particularly significant.}. We should also keep in
mind that experiments only provide information on the
Cabibbo--Kobayashi--Maskawa (CKM) quark mixing matrix
$V^{CKM}=V_{u}^{L\dagger}V_{d}^{L}$, and that one can construct
theoretical models where the left quark mixing is either in the up
sector, or in the down sector, or both. Similarly, the
Maki--Nakagawa--Sakata lepton mixing comes from the product of those
of charged leptons and neutrinos; with the additional complication
that, for the latter, we have both Dirac and Majorana mass terms.

Without entering into detailed model building, we generically observe
the following: 
\begin{itemize}
\item[(i)]
The right-handed quark mixing (relevant for $\bar{U}$ and $\bar{D}$)
is essentially not constrained by the data. Therefore, for a model
with left-right asymmetric mass matrices, one could also imagine a
manifestation with minimal mixing in the right-handed sector, in which
case a $\bar{U}_{3}\bar{D}_i\bar{D}_j$ operator would be the only
relevant one, and the gravitino would be essentially stable!
\item[(ii)]
For the left quark mixing (relevant for $Q$), we know the values from
$V_{CKM}$, where for instance the 2-3 mixing is a factor of $\approx
0.04$. Thus, a coupling $\lambda^{\prime} L_3 Q_3 \bar{D}_3$ also
implies the coupling $0.04 \lambda^{\prime} L_3 Q_2
\bar{D}_3$.
\item[(iii)]
The left lepton mixing (relevant for $L$) is constrained by the lepton
data, giving large 1-2 and 2-3 mixing, and small 1-3.
\end{itemize}

We see that there are in principle several flavour choices that can
lead to significant effects from the decays under discussion, and one
could in fact reverse the argument: the study of the relative
magnitude of radiative versus tree-level violating decays, may yield
relevant information for the flavour structure of the underlying
theory.

%%%%%%%%%%%%%%%%%%%%%%%%%%%%%%%%%%%%%%%%%%%%%%%%%%%%%%%%%%%%%%%%%%%%%%%%
\section{Radiative versus three-body decays}
\label{sec:radvs3b}
\setcounter{equation}{0}
%%%%%%%%%%%%%%%%%%%%%%%%%%%%%%%%%%%%%%%%%%%%%%%%%%%%%%%%%%%%%%%%%%%%%%%%

The decay rate for the radiative decay $\tilde G\to\gamma\nu$ is given by
\begin{equation}
\Gamma=\frac{1}{16\pi}\frac{1}{m_\gravitino}\overline{|{\cal M}|^2},
\end{equation}
where $\overline{|{\cal M}|^2}$ denotes the absolute square of the
Feynman amplitude, averaged over gravitino spin states, and summed
over photon polarisations. In terms of the decomposition
(\ref{Eq:calM}), the decay rate can be given by
\begin{equation}
\Gamma=\frac{\alpha\lambda^2\,m_\gravitino}{2048\pi^4}\,
\frac{m_\ell^2}{M^2}
\overline{\left|{\cal F}\right|^2},
\end{equation}
with
\begin{align}
\overline{\left|{\cal F}\right|^2}
&=\frac{1}{4}\sum_\text{spin, pol.}
\big|\bar u(p-k)P_R
[c_1{\cal A}_\mu^{(1)}+c_2{\cal A}_\mu^{(2)}]\tilde\psi^\mu(p)\big|^2
\nonumber \\
&=\frac{3}{4}|c_1|^2+\frac{2}{6}|c_2|^2-\frac{1}{6}\Re(c_1^*c_2),
\end{align}
where the averaging over gravitino spin states involves
\begin{align}
P_{\mu\nu}(p) 
\equiv &\sum_\text{spin} \tilde\psi_\mu(p)\overline{\tilde\psi}_\nu(p) 
\nonumber \\
=&-(\pslash-m_\gravitino)
\left\{\left(g_{\mu\nu}-\frac{p_\mu p_\nu}{m_\gravitino^2}\right)
-\frac{1}{3}
\left(g_{\mu\alpha}-\frac{p_\mu p_\alpha}{m_\gravitino^2}\right)
\left(g_{\nu\beta}-\frac{p_\nu p_\beta}{m_\gravitino^2}\right)
\gamma^\alpha\gamma^\beta\right\}.
\end{align}
The diagrams with reversed fermion flow, and right-handed s(fermions)
in the loop are found to have the same structure, differing only in
the sfermion loop mass. The decay rate for $\tilde G\to\gamma\bar\nu$
is identical, and will provide a factor of two in the total rate.

In Fig.~\ref{Fig:lifetime}, we compare the resulting lifetime of the
gravitino for radiative versus tree-level decays resulting from
trilinear R-violating terms. We fix $m_{\tilde\ell} = 200$~GeV and
$\lambda=0.001$. The interested reader should rescale the results to
suit his or her preferred value of $\lambda$, using $\tau\propto
1/\lambda^2$. The black lines denote radiative decays for (s)tau and
(s)muon loops, while the green lines represent the results for the
corresponding three-body decays, with intermediate stau and smuon,
taken from~\cite{CM}.

%%%%%%%%%%%%%%%%%%%%%%%%%%%%%%%%%%%%%%%%%%%%%%%%%%%%%%%%%%%%%%%%%%%%%%%%
\begin{figure}[htb]
\refstepcounter{figure}
\label{Fig:lifetime}
\addtocounter{figure}{-1}
\begin{center}
\setlength{\unitlength}{1cm}
\begin{picture}(15.0,7.5)
\put(1.5,0.)
{\mbox{\epsfysize=7.5cm
 \epsffile{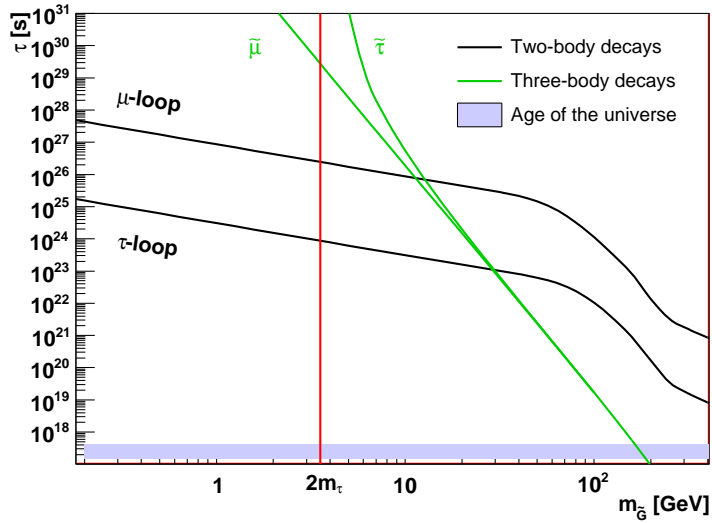}}}
\end{picture}
%\vspace*{-4mm}
\caption{Gravitino lifetime versus mass for 
two-body (black) and three-body decays (green). Also shown is
the area excluded by a universe age of $13.7$~Gyr (blue), and the
kinematical threshold for the two-tau final state of the three-body
decay (red line).}
\end{center}
\end{figure}
%%%%%%%%%%%%%%%%%%%%%%%%%%%%%%%%%%%%%%%%%%%%%%%%%%%%%%%%%%%%%%%%%%%%%%

We see that the radiative decays can easily dominate for gravitino
masses below $30$~GeV. While the three-body decay involving an
intermediate stau hits the kinematical threshold at $2m_\tau$, the
radiative dominance is still present for decays involving an
intermediate smuon, where there is no threshold at the mass scales
shown. For a selectron intermediary (not shown) the conclusions are
the same, with radiative dominance below a gravitino mass of around
$2$~GeV. This behaviour is controlled by the mass dependence of the
decay width: for the three-body decay $\Gamma_{\tilde G}\propto
m_{\tilde G}^7$, while for the radiative decay $\Gamma_{\tilde
G}\propto m_{\tilde G}$ at low gravitino masses. This can be
understood from Eq.~(\ref{eq:c1c2}), where we see that the leading
term in $c_1$ and $c_2$, for the limit of slepton masses much larger than the
gravitino and lepton masses, $2m_{\tilde l}^2C_0$, is
essentially independent of mass. The physical interpretation of this
is that the gravitational coupling compensates for the high loop-mass
by its increasing strength for higher masses. Because of the helicity
structure of the couplings, the two-body decay width is also heavily
dependent on the fermion mass, $\propto m_l^2$ at low gravitino
masses. Thus dominant third generation couplings, which imply a tau or
bottom quark, give significantly shorter lifetimes for the radiative
decay.

To constitute a realistic dark matter candidate the gravitino lifetime
should exceed the age of the universe. This lower boundary is shown as
a blue area in Fig.~\ref{Fig:lifetime}. For the R-parity violating couplings
considered here this only allows us to exclude gravitino masses above
$\sim 150$~GeV. However, we shall see in the next Section that the
photon flux from the gravitino decays can provide us with a much
stronger bound.

In Fig.~\ref{Fig:mgms}, we show the dependence of the gravitino
lifetime on the stau mass for the sum of radiative and three-body
decays, and again for a fixed $\lambda_{233}=0.001$. As expected the
gravitino mass where the three-body decay takes over depends on the
stau mass. The insensitivity of the lifetime to the slepton mass for
light gravitinos can again be understood from the dominant terms in
$c_1$ and $c_2$.

%%%%%%%%%%%%%%%%%%%%%%%%%%%%%%%%%%%%%%%%%%%%%%%%%%%%%%%%%%%%%%%%%%%%%%%%
\begin{figure}[htb]
\refstepcounter{figure}
\label{Fig:mgms}
\addtocounter{figure}{-1}
\begin{center}
\setlength{\unitlength}{1cm}
\begin{picture}(15.0,7.5)
\put(1.5,0.)
{\mbox{\epsfysize=8cm
 \epsffile{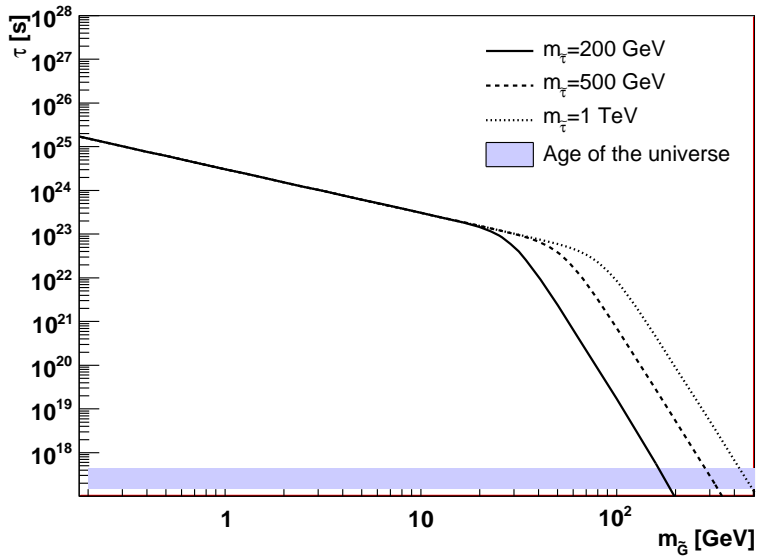}}}
\end{picture}
%\vspace*{-4mm}
\caption{Gravitino lifetime versus mass for different slepton masses.}
\end{center}
\end{figure}
%%%%%%%%%%%%%%%%%%%%%%%%%%%%%%%%%%%%%%%%%%%%%%%%%%%%%%%%%%%%%%%%%%%%%%

%%%%%%%%%%%%%%%%%%%%%%%%%%%%%%%%%%%%%%%%%%%%%%%%%%%%%%%%%%%%%%%%%%%%%%%%
\section{Cosmological implications and collider signatures}
\label{sec:cosmocoll}
\setcounter{equation}{0}
%%%%%%%%%%%%%%%%%%%%%%%%%%%%%%%%%%%%%%%%%%%%%%%%%%%%%%%%%%%%%%%%%%%%%%%%

From the above we see that the mass and lifetime of the gravitino in
principle allow it to be dark matter. Whether this could be the case
for a specific model depends on the predicted gravitino density. If
the universe has gone through a period of inflation, the primordial
gravitino abundance is erased, and gravitinos are created either
thermally or from the decays of the NLSP\footnote{With non-zero
R-parity violation the NLSP production mechanism no longer
applies.}. This implies that the abundance of gravitinos is very
sensitive to the reheating temperature of the universe --- which is an
open question --- and can well be in the correct range for gravitinos
to be dark matter, or a component of it.

In MSSM dark matter considerations, the allowed parameter space is
severely constrained by possible effects of late NLSP decays on the
light element abundances, as predicted by Big Bang Nucleosynthesis
\cite{MSSM-DM}. However, as has been remarked by several authors
\cite{TY,BM,Gherghetta:1998tq}, in models where the NLSP decays
via R-violating couplings, the BBN bounds are easily satisfied.

With regards to baryogenesis and leptogenesis it has been observed
that severe constraints on $R$-parity violating interactions can be
derived from cosmological arguments~\cite{CDEO}: a pre-existing baryon
asymmetry would be erased if the interactions are strong enough to
come to equilibrium at the time of the electroweak phase transition.
However, it has been argued that this need not be the case in various
schemes where flavour effects in leptogenesis/baryogenesis are
considered \cite{DRR}. In our analysis therefore we
consider larger R-violating couplings than those used in
e.g. \cite{BM}.

Our final cosmological consideration is whether the presence of
photons in the final state is compatible with measurements on the
extragalactic diffuse photon background, and may even be reconciled
with an apparent excess in the EGRET data, as proposed in \cite{BM}.
This will in fact give the strictest bounds on R-violating couplings,
for the case of radiative decays.

The photon flux from gravitino decays is described by a red-shifted
monochromatic line, which is given by \cite{BM}:
\begin{equation}
F(E)=E^2\frac{dJ}{dE}={\rm BR}(\tilde G\to\gamma\nu)
\times C_\gamma(1+\kappa x^3)^{-1/2}x^{5/2}\theta(1-x),
\end{equation}
where
\begin{equation}
x=\frac{2E}{m_{\tilde G}}, \quad
C_\gamma=\frac{\Omega_{\tilde G}\rho_c}
{8\pi\tau_{\tilde G\to\gamma\nu} H_0 \Omega_M^{1/2}}
\quad {\rm and} \quad \kappa=\frac{\Omega_\Lambda}{\Omega_M}.
\end{equation}
With current values for the cosmological parameters \cite{Yao:2006px}
we arrive at\footnote{Note that there is an order of magnitude
difference between our value, and that of \cite{BM}. We have learnt
from the authors that this is due to a misprint in their paper.}
\begin{equation}
C_\gamma = 1.06\cdot 10^{-6}\left(\frac{10^{27}{\rm s}}
{\tau_{\tilde G\to\gamma\nu}}\right)
{\rm GeV~cm^{-2}~sr^{-1}~s^{-1}} \quad {\rm and} \quad \kappa\simeq 3.
\end{equation}
For comparison, the original EGRET analysis \cite{EGRET} gave a power
law description of the extragalactic flux as
\begin{equation}
E^2\frac{dJ}{dE}=1.37\cdot 10^{-6}\left(\frac{1~{\rm GeV}}{E}\right)^{0.1}
{\rm GeV~cm^{-2}~sr^{-1}~s^{-1}},
\end{equation}
in the energy range 30~MeV to 100~GeV. The non-observation of an
excess with monochromatic origin can be used to restrict the gravitino
mass and lifetime.

In Fig.~\ref{Fig:max}, we show the maximum allowed value of the
R-violating coupling $\lambda$ for a range of gravitino
masses\footnote{Similar considerations are directly applicable to
$\lambda^\prime$.}. We require that the gravitinos can be dark matter,
with a lifetime of at least $10$ times the current age of the
universe, and that their radiative decays are consistent with the
photon spectrum measured by EGRET, extrapolating the power law
behaviour up to 400~GeV. The result is a simple linear dependence on
the log of the gravitino mass for the range where the radiative decays
dominate. For higher gravitino masses, where the three-body decays
become important, the bound relaxes as the branching ratio to photons
decreases dramatically. Eventually the bound on the lifetime, 
for gravitinos to be dark matter,
becomes dominant and the maximum allowed value decreases rapidly
towards the slepton mass threshold. The extra features seen at the
very highest masses are due to threshold effects for the three-body
decays.

%%%%%%%%%%%%%%%%%%%%%%%%%%%%%%%%%%%%%%%%%%%%%%%%%%%%%%%%%%%%%%%%%%%%%%%%
\begin{figure}[htb]
\refstepcounter{figure}
\label{Fig:max}
\addtocounter{figure}{-1}
\begin{center}
\setlength{\unitlength}{1cm}
\begin{picture}(15.0,7.5)
\put(1.5,0.)
{\mbox{\epsfysize=7.5cm
 \epsffile{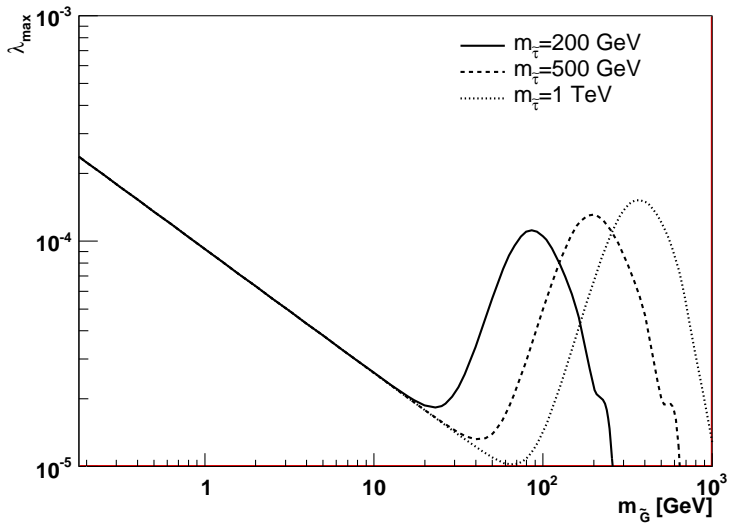}}}
\end{picture}
%\vspace*{-4mm}
\caption{Maximum value of R-violating coupling $\lambda$ versus
gravitino mass.}
\end{center}
\end{figure}
%%%%%%%%%%%%%%%%%%%%%%%%%%%%%%%%%%%%%%%%%%%%%%%%%%%%%%%%%%%%%%%%%%%%%%

We find that the allowed couplings are small for radiative decays in
an intermediate mass range, but can be significantly larger when
tree-level decays dominate or for very light gravitino masses. In all
cases the couplings can be sufficiently large to lead to interesting
expectations for collider phenomenology.

It is interesting to note that the strict upper bounds on the
couplings derived from the photon spectrum, naturally bring us to the
range that is relevant for the generation of neutrino masses
\cite{Hall:1983id,constraints,neut-masses} compatible with neutrino data.
Moreover, the power law excess claimed for the 2-10~GeV photon energy
range in the more recent analysis of the EGRET data \cite{reEGRET}
could easily be explained by decaying gravitino dark matter with a
mass of around 10~GeV and R-violating couplings of ${\cal
O}(10^{-4})$.

Moving on to the collider signatures, these will be the standard
signatures of R-violating supersymmetry, with multi-lepton or
multi-jet events and the possibility of explicit lepton number
violation at the final state. Given the magnitude of R-violating
couplings required from the photon spectrum and for gravitino dark
matter, MSSM production of sparticle pairs followed by R-violating
decays would be expected. However, for the parameter space and flavour
structure where tree-body decays dominate, single superparticle
productions may still occur along the lines previously proposed in the
literature
\cite{Rpar,Hall:1983id,DR}. 

The specific decay modes depend on the NLSP and the branching ratios
of R-violating versus R-conserving modes. If the NLSP is the
neutralino it will decay to three fermions, with diagrams similar to
those of the tree-level decay of the gravitino, see
Fig.~\ref{Fig:3-body}. If instead the NLSP is a slepton ---
e.g. $\tilde{\nu}$ or even $\tilde\tau_R$ --- its decay will
dominantly proceed through a direct decay to two fermions, for the
cases with large R-violating couplings involving its flavour, or
through an intermediary neutralino to four fermions (for small
R-violating couplings).  The specific branching ratios depend on the
supersymmetric parameter space and the magnitude of the R-violating
coupling, and have been extensively studied
\cite{Rpar,Hall:1983id,DR,AEGLM}.

\begin{table}[!ht]
\begin{center}
\begin{tabular}{|c|c|c|c|}
\hline
\phantom{\Big|}
NLSP 
& $LL\bar{E}$ 
& $LQ{\bar{D}}$ 
& $\bar{U}\bar{D}\bar{D}$  
\\ \hline
$ \tilde\chi_1^0 $ 
& $\ell_i^{\pm} {\ell}_j^{\mp} \nu$ 
& $q_j \bar{q}_k\ell^{\pm}(q_j \bar{q}_k \nu)$
& $q_iq_jq_k (\bar{q}_i\bar{q}_j\bar{q}_k)$
\\ \hline
$ \tilde{\nu}$ 
& $\ell_i^{\pm} {\ell}_j^{\mp}$ 
& $q_j \bar{q}_k$
&          
\\  
& $\ell_i^{\pm} {\ell}_j^{\mp} \nu \nu$ 
& $q_j \bar{q}_k\ell^{\pm}\nu (q_j \bar{q}_k \nu \nu)$
& $\nu q_iq_jq_k (\nu \bar{q}_i\bar{q}_j\bar{q}_k)$
\\ \hline 
$ \tilde\tau_R$ 
& $\ell_i \nu$ 
& $q_j \bar{q}_k$
&          
\\  
& $\ell_i^{\pm} {\ell}_j^{\mp} \nu \tau$ 
& $q_j \bar{q}_k\ell^{\pm}\tau (q_j \bar{q}_k \nu \tau)$
& $\tau q_iq_jq_k (\tau \bar{q}_i\bar{q}_j\bar{q}_k)$
\\ \hline 
\end{tabular}
\caption{\it NSLP R-violating decays.}
\end{center}
\label{NLSP}
\end{table}

The choice of NLSP in the above table has been made according to the
principle that, for universal sparticle masses at a high scale, the
running of couplings to low energies implies an increase in masses due
to gauge interactions and a decrease due to Yukawa couplings. Thus,
the neutralino and the sneutrino, having no electric charge, and the
right-handed stau, due to the lack of weak interactions and a large
Yukawa coupling in supersymmetric models with large $\tan\beta$, are
the most obvious candidates.

%%%%%%%%%%%%%%%%%%%%%%%%%%%%%%%%%%%%%%%%%%%%%%%%%%%%%%%%%%%%%%%%%%%%%%%%
\section{Conclusions}
\setcounter{equation}{0}
%%%%%%%%%%%%%%%%%%%%%%%%%%%%%%%%%%%%%%%%%%%%%%%%%%%%%%%%%%%%%%%%%%%%%%%%

We have studied radiative gravitino decays to a photon and a neutrino,
generated from trilinear R-violating operators. We calculated the
decay rate, and compared it to tree-level decays of the gravitino to
three fermions, which arise from the same operators. There is a wide
region of parameters for which the loop suppression of the radiative
mode is less effective than the phase-space suppression of the
tree-level one, particularly for light gravitino masses and
R-violating operators that involve the third generation of fermions.

Whether enhanced gravitino decays can be expected, depends
on the flavour structure of the R-violating
operators.  Without entering into details of model building, it is
nevertheless possible to make simple analogies, and link R-violating
couplings to those that generate Yukawa mass terms for the
fermions. Mixing effects can also be important, since tree-level
decays that are disallowed by phase-space considerations for certain
operator flavours could nevertheless proceed through quark mixing
effects, suppressed by the mixing factors. For a dominant operator
$\bar{U}_3\bar{D}_i\bar{D}_j$, the absence of the radiative mode and
the top quark in the vertex, imply that the gravitino can be
essentially stable in models with small right-handed quark mixing.

We find that the gravitino lifetime is typically longer than the age
of the universe due to suppression by Planck scale and loop effects,
meaning that gravitinos can constitute dark matter. The R-violating
couplings and sparticle masses in such scenarios have values that can
give rise to distinct signals in colliders, and the R-violating
couplings are in general higher than those expected in bilinear
R-violating schemes \cite{BM}.

However, for the parameter space where radiative decays dominate,
photon spectra further constrain the magnitude of R-violating
couplings, particularly for $L_{1,2}L_3\bar{E}_3$ and
$L_{1,2,3}Q_3\bar{D}_3$ operators. In fact the presence of photons in
the final state of R-violating gravitino decays is compatible with an
excess seen in the EGRET data on extragalactic diffuse gamma-rays, as
proposed in \cite{BM}. Moreover, the range of couplings allowed by
these considerations is compatible with the one required in order to
generate acceptable neutrino mass patterns.

Finally, the decays of the next-to-lightest supersymmetric particle
can still proceed via the R-violating vertexes at significant rates,
and are thus easily accommodated within the framework of Big Bang
Nucleosynthesis.

\vspace*{0.2 cm}

%%%%%%%%%
{\bf Acknowledgements.}  It is a pleasure to thank W. Buchm\"uller for
input on the gravitino Feynman rules.  The research of S. Lola
is co-funded by the FP6 Marie Curie Excellence Grant
MEXT-CT-2004-014297.  Additional support for research visits has been
provided by the European Network MRTPN-CT-2006 035863-1 (UniverseNet).
The research of P. Osland and A. R. Raklev has been supported by the
Research Council of Norway.  All authors are grateful to the
organisers of the CERN Theory Institute on LHC-Cosmology Interplay for
hospitality during the last stages of this project.

%%%%%%%%%%%%%%%%%%%%%%%%%%%%%%%%%%%%%%%%%%%%%%%%%%%%%%%%%%%%%%%%%%%%%%

\end{document}